# Relation Strength-Aware Clustering of Heterogeneous Information Networks with Incomplete Attributes[*]

Yizhou Sun[†]
University of Illinois at
Urbana-Champaign
Urbana, IL
sun22@illinois.edu

Charu C. Aggarwal
IBM T. J. Watson Research
Center
Yorktown Heights, NY
charu@us.ibm.com

Jiawei Han
University of Illinois at
Urbana-Champaign
Urbana, IL
hanj@cs.uiuc.edu

## ABSTRACT

With the rapid development of online social media, online shopping sites and cyber-physical systems, heterogeneous information networks have become increasingly popular and content-rich over time. In many cases, such networks contain multiple types of objects and links, as well as different kinds of attributes. The clustering of these objects can provide useful insights in many applications. However, the clustering of such networks can be challenging since (a) the attribute values of objects are often incomplete, which implies that an object may carry only partial attributes or even no attributes to correctly label itself; and (b) the links of different types may carry different kinds of semantic meanings, and it is a difficult task to determine the nature of their relative importance in helping the clustering for a given purpose. In this paper, we address these challenges by proposing a model-based clustering algorithm. We design a probabilistic model which clusters the objects of different types into a common hidden space, by using a user-specified set of attributes, as well as the links from different relations. The strengths of different types of links are automatically learned, and are determined by the given purpose of clustering. An iterative algorithm is designed for solving the clustering problem, in which the strengths of different types of links and the quality of clustering results mutually enhance each other. Our experimental results on real and synthetic data sets demonstrate the effectiveness and efficiency of the algorithm.

## 1. INTRODUCTION

With the rapid emergence of online social media, online shopping sites and cyber-physical systems, it has become possible to model many forms of interconnected networks as heterogeneous information networks in which objects (*i.e.*, nodes) are of different types, and links among objects correspond to different relations, denoting different interaction semantics. An object is usually associated with some attributes. For example, in the case of the *YouTube* social media network, the object types include videos, users, and comments; links between objects correspond to different relations, such as publish and like relations between users and videos, post relation between users and comments, friendship and subscribe relations between users, and so on; and attributes include user's location, video's clip length and number of views, comments, and so on.

Such kinds of heterogeneous information networks are ubiquitous and the determination of their underlying clusters has many interesting applications. For example, clustering objects (customers, products, comments, etc.) in an online shopping network such as *eBay* is helpful for customer segmentation in product marketing; and clustering objects (people, groups, books, posts, etc.) in an online social network such as *Facebook* is helpful for voter segmentation in political campaigns. Another example is the weather sensor network, where different types of sensors may carry different numerical attributes and be linked by $k$ nearest neighbor relationships. The clustering process may reveal useful regional weather patterns.

The clustering task brings two new challenges in such scenarios. First, an object may contain only partial or even no observations for a given attribute set that is critical to determine their cluster labels. That is, a pure attribute-based clustering algorithm cannot correctly detect these clusters. Second, although links have been frequently used in networks to detect clusters [8, 17, 1, 23] in recent research, we consider a much more challenging scenario in which the links are of different types and interpretations, each of which may have its own level of semantic importance in the clustering process. That is, a pure link-based clustering without any guidance from attribute specification could fail to meet user demands.

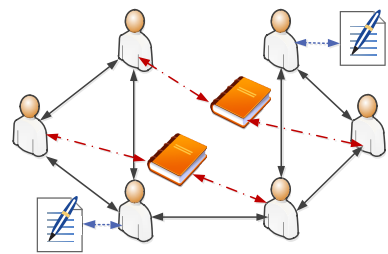

Figure 1: A Motivating Example on Clustering Political Interests in Social Information Networks

[*]The work was supported in part by the U.S. Army Research Laboratory under Cooperative Agreement No. W911NF-09-2-0053 (NS-CTA), MIAS, a DHS-IDS Center for Multimodal Information Access and Synthesis at UIUC, U.S. National Science Foundation grants IIS-0905215, and U.S. Air Force Office of Scientific Research MURI award FA9550-08-1-0265.

[†]The work was partially performed during Yizhou Sun's employment at IBM T. J. Watson Research Center.

Fig. 1 shows a toy social information network extracted from a political forum containing users, blogs written by users, books liked by users, and friendship between users. Now suppose we want to cluster users in the network according to their political interests, using the text attributes in user profiles, blogs and books, as well as the link information between objects. On one hand, since not all the users listed their political interests in their profiles, we cannot judge their political interests simply according to the text information contained in their profiles directly. On the other hand, without specifying the purpose of clustering, we cannot decide which types of links to use for the clustering: Shall we use the friendship links to detect the social communities, or the user-like-book links to detect the reading groups, or a mix of them? Obviously, to solve such clustering tasks, we need to use both the incomplete attribute information as well as the link information of different types with the awareness of their importance weights. In our example, in order to discover a user's political interests, we need to learn which link types are more important for our purpose of clustering, among the relationships between her and blogs, books, and her friends.

Recently, there have been several studies [28, 18, 20, 25, 24, 16] showing that the combination of attribute and link information in a network can improve the clustering quality. However, none of these studies has addressed the two challenges simultaneously. Many of the studies [28, 20, 25] rely on a complete attribute space and the clustering result is considered as a trade-off between attribute-based measures and link-based measures. Moreover, none of the current studies has examined the issue that different types of links have different importance in determining a clustering with a certain purpose.

In this study, we explore the interplay between different types of links and the specified attribute set in clustering process, and design a comprehensive and robust probabilistic clustering model for heterogeneous information networks. First, we model each attribute attached with each object as a mixture model, with the mixing proportion as the soft clustering probability for each object. As it is a generative model, the incompleteness issue of the attributes is handled properly. Second, the importance of different types of links is modeled with different coefficients, which is determined by the consistency of cluster membership vectors over all the linked objects. In other words, the cluster membership information of objects are propagating in the whole network, but different types of links carry different capabilities in the propagation process. The goal is to determine the optimal levels of importance of the different types of semantic links and the clustering results for objects simultaneously. An iterative method is proposed to learn the parameters, where the clustering results and the importance weights for different link types are optimized alternately and mutually enhance each other.

The primary contributions of this paper are as follows.

1. We propose a clustering problem for heterogeneous information networks with incomplete attributes across objects and different types of links, according to a user-specified attribute set that may be from different types.

2. We design a novel probabilistic clustering model, which for the first time directly models the varying importance of different types of semantic links, for the above clustering problem.

3. We propose an efficient algorithm to compute this model, where the clustering results and strengths for different typed links mutually enhance each other.

4. We present experiments on both real and synthetic data sets to demonstrate the effectiveness and efficiency of the method.

## 2. PROBLEM DEFINITION

In this section, we introduce the notations, definitions and concepts relevant to the problem of clustering heterogeneous networks.

### 2.1 The Data Structure

A **heterogeneous information network** $G = (V, E, W)$ is modeled as a directed graph, where each node $v \in V$ in the network corresponds to an object (or an event), and each link $e \in E$ corresponds to a relationship between the linked objects, with its weight denoted by $w(e)$. Different from the traditional network definition, the objects and links in heterogeneous networks are associated with explicit type information to distinguish the semantic meanings, namely, we have a mapping function from object to object type, $\tau : V \to \mathcal{A}$, and a mapping function from link to link type, $\phi : E \to \mathcal{R}$. $\mathcal{A}$ is the object type set, and $\mathcal{R}$ is the link type set, or the relation set, which provides linkage guidance between nodes. Notice that, if a relation exists from type $A$ to type $B$, denoted as $A\,R\,B$, the inverse relation $R^{-1}$ holds naturally for $B\,R^{-1}\,A$. For most of the times, $R$ and its inverse $R^{-1}$ are not equal, unless the two types are the same and $R$ is symmetric.

**Attributes** are associated with objects, such as the location of a user, the text description of a book, the text information of a blog, and so on. In this setting, we consider attributes across all different types of objects as a collection of attributes for the network, denoted as $\mathcal{X} = \{X_1, \ldots, X_T\}$, in which we are interested only in a subset for a certain clustering purpose. Each object $v \in V$ contains a subset of the attributes, with **observations** denoted as $v[X] = \{x_{v,1}, x_{v,2}, \ldots, x_{v,N_{X,v}}\}$, where $N_{X,v}$ is the total number of observations of attribute $X$ attached with object $v$. Notice that, some attributes can be shared by different types of objects, such as the text and the location attribute; while some other attributes are unique for a certain type of objects, such as the clip time length for a video. We use $V_X$ to denote the object set that contains attribute $X$.

### 2.2 The Clustering Problem

In this paper, we study the clustering problem that maps every object in the network into a unified hidden space, *i.e.*, a soft clustering, according to the user-specified subset of attributes in the network, with the help of links from different types.

There are several new challenges for clustering objects in this new scenario. First, the attributes are usually **incomplete** for an object: the attributes specified by a user may be only partially or even not contained in an object type; and the values for these attributes could be missing even if the attribute type is contained in the object type. Moreover, the incompleteness of the data cannot be easily handled by interpolation: the observations for each attribute could be a set or a bag of values, and the neighbors for an object are from different types of objects, which may not be helpful for predicting the missing data. For example, it is impossible to get a user's blog via interpolating techniques. Therefore, none of the existing clustering algorithms that purely based on attribute space can solve the clustering problem in this scenario.

Second, with the awareness that links play a very important role to propagate the cluster information among objects, another challenge is that **different link types** have different semantic meanings and therefore have different strengths in the process of passing cluster information around. In other words, while it is clear that the existence of links between nodes is indicative of clustering similarity, it is also important to understand that *different link types may have a different level of importance in the clustering process.* In the example of clustering political interests illustrated in Fig. 1, we expect a higher importance of the relation *user-like-book* than the relation *friendship* in deciding the cluster membership of a user. Thus, we need to design a clustering model which can learn the importance of these link types automatically. This will enhance the clustering quality because it marginalizes the impact of low quality types of neighbors of an object during the clustering process.

We present examples of clustering tasks in two concrete heterogeneous information networks in the following.

EXAMPLE 1. ***Bibliographic information network.*** *A bibliographic network is a typical heterogeneous network, containing objects from three types of entities, namely papers, publication venues (conferences or journals), and authors. Each paper has different link types to its authors and publication venue. Each paper is associated with the text attribute as a bag of words. Each author and venue links to a set of papers, but contains no attributes (in our case). The application of a clustering process according to the text attribute in such a scenario can help detect research areas, and decide the research areas for authors, venues and papers.*

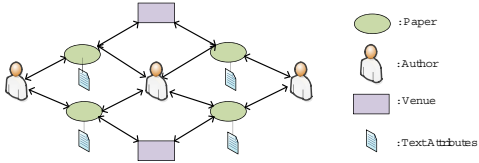

**Figure 2: Illustration of Bibliographic Information Network**

Multiple types of objects and links in this network are illustrated in Fig. 2. For objects of different types, their cluster memberships may need to be determined by different kinds of information: for authors and venues, the only available information is from the papers linked to them; for papers, both text attributes and links of different types are provided. Note that, even for papers that are associated with text attributes, using link information can further help the clustering quality when the observations of the text data is very limited (e.g., using text merely from titles). Also, we may expect that the neighbors of an author type play a more important role in deciding a paper's cluster compared with the neighbor of a venue type. This needs to be automatically learned in terms of the underlying relation strengths.

EXAMPLE 2. ***Weather sensor network.*** *Weather sensor networks typically contain different kinds of sensors for detecting different attributes, such as precipitation or temperature. Some sensors may have incorrect or no readings because of the inaccuracy or malfunctioning of the instruments. The links between sensors are generated according to their k nearest neighbors under geo-distances, in order to incorporate the importance of locality in weather patterns. The clustering of such sensors according to both precipitation and temperature attributes can be useful in determining regional weather patterns.*

Fig. 3 illustrates a weather sensor network containing two types of sensors: temperature and precipitation. A sensor may sometimes register none or multiple observations. Although it is desirable to use the complete observations on both temperature and precipitation to determine the weather pattern of a location, in reality a sensor object may contain only partial attribute (e.g., temperature values only for temperature sensors), and both the attribute and link information are needed for correctly detecting the clusters. Still, which type of links plays a more important role needs to be determined in the clustering process.

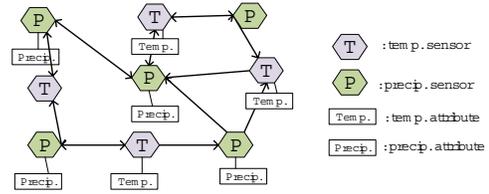

**Figure 3: Illustration of Weather Sensor Information Network**

Formally, given a network $G = (V, E, W)$, a specified subset of its associated attributes $X \in \mathcal{X}$, the attribute observations $\{v[X]\}$ for all objects, and the number of clusters $K$, our goal is:

1. to learn a soft clustering for all the objects $v \in V$, denoted by a membership probability matrix, $\Theta_{|V| \times K} = (\boldsymbol{\theta}_v)_{v \in V}$, where $\Theta(v, k)$ denotes the probability of object $v$ in cluster $k$, $0 \leq \Theta(v, k) \leq 1$ and $\sum_{k=1}^{K} \Theta(v, k) = 1$, and $\boldsymbol{\theta}_v$ is the $K$ dimensional cluster membership vector for object $v$, and

2. to learn the strengths (importance weights) of different link types in determining the cluster memberships of the objects, $\boldsymbol{\gamma}_{|\mathcal{R}| \times 1}$, where $\boldsymbol{\gamma}(r)$ is a real number and stands for the importance weight for the link type $r \in \mathcal{R}$.

Note that, in this paper we will not study the problem of how to determine the best number of clusters $K$, which belongs to the model selection problem and has been covered in a large number of studies by using various criteria [19, 12], such as AIC and BIC for probabilistic models.

## 3. THE CLUSTERING MODEL

We propose a novel probabilistic clustering model in this section and introduce the algorithm that optimizes the model in Section 4.

### 3.1 Model Overview

Given a network $G$, with the observations of its links and the observations $\{v[X]\}$ for the specified attributes $X \in \mathcal{X}$, a good clustering configuration $\Theta$, which can be viewed as hidden cluster information for objects, should satisfy two properties:

1. Given the clustering configuration, the observed attributes should be generated with a high probability. Especially, we model each attribute for each object as a separate mixture model, with each component representing a cluster.

2. The clustering configuration should be highly consistent with the network structure. In other words, linked objects should have similar cluster membership probabilities, and larger strength of a link type requires more similarity between the linked objects of this type.

Overall, we can define the likelihood of the observations of all the attributes $X \in \mathcal{X}$ as well as the hidden continuous cluster configuration $\Theta$, given the underneath network $G$, the relation strength vector $\boldsymbol{\gamma}$, and the cluster component parameter $\boldsymbol{\beta}$, which can be decomposed into two parts, the generative probability of the observed attributes given $\Theta$ and the probability of $\Theta$ given the network structure:

$$p(\{\{v[X]\}_{v \in V_X}\}_{X \in \mathcal{X}}, \Theta | G, \boldsymbol{\gamma}, \boldsymbol{\beta})$$
$$= \prod_{X \in \mathcal{X}} p(\{v[X]\}_{v \in V_X} | \Theta, \boldsymbol{\beta}) p(\Theta | G, \boldsymbol{\gamma}) \quad (1)$$

From a generative point of view, this model explains how observations for attributes associated with objects are generated: first,

a hidden layer of variables $\Theta$ is generated according to the probability $p(\Theta|G, \gamma)$, given the network structure $G$ and the strength vector $\gamma$; second, the observed values of attributes associated with each object are generated according to mixture models, given the cluster membership of the object, as well as the cluster component parameter $\beta$, with the probability $\prod_{X \in \mathcal{X}} p(\{v[X]\}_{v \in V_X}|\Theta, \beta)$.

The goal is then to find the best parameters $\gamma$ and $\beta$, as well as the best clustering configuration $\Theta$ that maximize the likelihood. The detailed modeling of the two parts is introduced in the following.

## 3.2 Modeling Attribute Generation

Given a configuration $\Theta$ for the network $G$, namely, the membership probability vector $\theta_v$ for each object $v$, the attribute observations for each object $v$ are conditionally independent with observations from other objects. Each attribute $X$ associated with each object $v$ is then assumed following the same family of mixture models that share the same cluster components, with the component mixing proportion as the cluster membership vector $\theta_v$. For simplicity, we first assume that only one attribute $X$ is specified for the clustering purpose and then briefly discuss a straightforward extension to the multi-attribute case.

### 3.2.1 Single Attribute

Let $X$ be the only attribute we are interested in the network, and let $v[X]$ be the observed values for object $v$, which may contain multiple observations. It is natural to consider that the attribute observation $v[X]$ for each object $v$ is generated from a mixture model, where each component is a probabilistic model that stands for a cluster, with the parameters to be learned, and component weights denoted by $\theta_v$. Formally, the probability of all the observations $\{v[X]\}_{v \in V_X}$ given the network configuration $\Theta$ is modeled as:

$$p(\{v[X]\}_{v \in V_X}|\Theta, \beta) = \prod_{v \in V_X} \prod_{x \in v[X]} \sum_{k=1}^{K} \theta_{v,k} p(x|\beta_k) \quad (2)$$

where $K$ is the number of clusters, and $\beta_k$ is the parameter for component $k$. In this paper, we consider two types of attributes, one corresponding to text attributes with categorical distributions, and the other numerical attributes with Gaussian distributions.

**(1) Text attribute with categorical distribution:** In this case, objects in the network contain text attributes in the form of a term list, from the vocabulary $l = 1$ to $m$. Each cluster $k$ has a different term distribution following a categorical distribution, with the parameter $\beta_k = (\beta_{k,1}, \ldots, \beta_{k,m})$, where $\beta_{k,l}$ is the probability of term $l$ appearing in cluster $k$, i.e., $X|k \sim discrete(\beta_{k,1}, \ldots, \beta_{k,m})$. Following the frequently used topic modeling method PLSA [11], each term in the term list for an object $v$ is generated from the mixture model, with each component as a categorical distribution over terms described by $\beta_k$, and the component coefficient is $\theta_v$. Formally, the probability of observing all the current attribute values is:

$$p(\{v[X]\}_{v \in V_X}|\Theta, \beta) = \prod_{v \in V_X} \prod_{l=1}^{m} (\sum_{k=1}^{K} \theta_{v,k} \beta_{k,l})^{c_{v,l}} \quad (3)$$

where $c_{v,l}$ denotes the count of term $l$ that object $v$ contains.

**(2) Numerical attribute with Gaussian distribution:** In this case, objects in the network contain numerical observations in the form of a value list, from the domain $\mathbb{R}$. The $k$th cluster is a Gaussian distribution with parameters $\beta_k = (\mu_k, \sigma_k^2)$, i.e., $X|k \sim \mathcal{N}(\mu_k, \sigma_k^2)$, where $\mu_k$ and $\sigma_k$ are mean and standard deviation of normal distribution for component $k$. Each observation in the observation list for an object $v$ is generated from the Gaussian mixture model, where each component is a Gaussian distribution with parameters $\mu_k, \sigma_k^2$, and the component coefficient is $\theta_v$. The probability density for all the observations for all objects is then:

$$p(\{v[X]\}_{v \in V_X}|\Theta, \beta) = \prod_{v \in V_X} \prod_{x \in v[X]} \sum_{k=1}^{K} \theta_{v,k} \frac{1}{\sqrt{2\pi\sigma_k^2}} e^{-\frac{(x-\mu_k)^2}{2\sigma_k^2}} \quad (4)$$

### 3.2.2 Multiple Attributes

As in the weather sensor network example, we are interested in multiple attributes, namely temperature and precipitation. Generally, if multiple attributes in the network are specified by users, say $X_1, \ldots, X_T$, the probability density of observed attribute values $\{v[X_1]\}, \ldots, \{v[X_T]\}$ for a given clustering configuration $\Theta$ is as follows, by assuming the independence among these attributes:

$$\begin{aligned} &p(\{v[X_1]\}_{v \in V_{X_1}}, \ldots, \{v[X_T]\}_{v \in V_{X_T}}|\Theta, \beta_1, \ldots, \beta_T) \\ &= \prod_{t=1}^{T} p(\{v[X_t]\}_{v \in V_{X_t}}|\Theta, \beta_t) \end{aligned} \quad (5)$$

## 3.3 Modeling Structural Consistency

From the view of links, the more similar the two objects are in terms of cluster membership, the more likely they are connected by a link. In order to quantitatively measure the consistency of a clustering result $\Theta$ with the network structure $G$, we define a novel probability density function for observing $\Theta$.

We assume that linked objects are more likely to be in the same cluster, if the link type is of importance in determining the clustering process. That is, for two linked objects $v_i$ and $v_j$, their membership probability vectors $\theta_i$ and $\theta_j$ should be similar. Within the same type of links, the higher link weight ($w(e)$), the more similar $\theta_i$ and $\theta_j$ should be. Further, a certain link type may be of greater importance, and will influence the similarity to a greater extent. The consistency of a configuration $\Theta$ with the network $G$, is evaluated with the use of a composite analysis with respect to all the links in the network in the form of a probability density value. A more consistent configuration of $\Theta$ will yield a higher probability density value. In the following, we first introduce how the consistency of two cluster membership vectors is defined with respect to a single link, and then how this analysis can be applied over all links in order to create a probability density value as a function of $\Theta$.

For a link $e = \langle v_i, v_j \rangle \in E$, with type $r = \phi(e) \in \mathcal{R}$, we denote the *importance of the link type to the clustering process* by a real number $\gamma(r)$. This is different from the weight of the link $w(e)$, which is specified in the network as input, whereas the value of $\gamma(r)$ is defined on link types and needs to be learned. We denote the consistency function of two cluster membership vectors $\theta_i$ and $\theta_j$ with link $e$ under strength weights for each link type $\gamma$ by a *feature function* $f(\theta_i, \theta_j, e, \gamma)$. Higher values of this function imply greater consistency with the clustering results. In the following, we list several desiderata for a good feature function:

1. The value of the feature function $f$ should increase with greater similarity of $\theta_i$ and $\theta_j$.

2. The value of the feature function $f$ should decrease with greater importance of the link $e$, either in terms of its specified weight $w(e)$, or learned importance $\gamma(r)$. In other words, for the larger strength of a particular link type, two linked nodes are required to be more similar to claim the same level of consistency.

3. The feature function should not be symmetric between its first two arguments $\theta_i$ and $\theta_j$, because the impact from node $v_i$ to node $v_j$ could be different from that of $v_j$ to $v_i$.

The last criterion requires some further explanation. For example, in a citation network, a paper $i$ may cite paper $j$, because $i$ feels that $j$ is relevant to itself, while the reverse may not be necessarily true. In the experimental section, we will show that asymmetric feature functions produce higher accuracy in link prediction.

We then propose a cross entropy-based feature function, which satisfies all of the desiderata listed above. For a link $e = \langle v_i, v_j \rangle \in E$, with relation type $r = \phi(e) \in \mathcal{R}$, the feature function $f(\boldsymbol{\theta}_i, \boldsymbol{\theta}_j, e, \boldsymbol{\gamma})$ is defined as:

$$f(\boldsymbol{\theta}_i, \boldsymbol{\theta}_j, e, \boldsymbol{\gamma}) = -\gamma(r) w(e) H(\boldsymbol{\theta}_j, \boldsymbol{\theta}_i) = \gamma(r) w(e) \sum_{k=1}^{K} \theta_{j,k} \log \theta_{i,k} \quad (6)$$

where $H(\boldsymbol{\theta}_j, \boldsymbol{\theta}_i) = -\sum_{k=1}^{K} \theta_{j,k} \log \theta_{i,k}$, is the cross entropy from $\boldsymbol{\theta}_j$ to $\boldsymbol{\theta}_i$, which evaluates the deviation of $v_j$ from $v_i$, in terms of the average coding bits needed if using coding schema based on the distribution of $\boldsymbol{\theta}_i$. For a fixed value of $\gamma(r)$, the value of $H(\boldsymbol{\theta}_j, \boldsymbol{\theta}_i)$ is minimal and (therefore) $f$ is maximal, when the two vectors are identical. It is also evident from Eq. (6) that the value of $f$ decreases with increasing learned link type strength $\gamma(r)$ or input link weight $w(e)$. We require $\boldsymbol{\gamma} \geq 0$, in the sense that we do not consider links that connect dissimilar objects. The value of $f$ so defined is a non-positive function, with larger value indicating a higher consistency of the link.

Other distance functions such as KL-divergence could replace the cross entropy in the feature function. However, as cross entropy favors distributions that concentrate on one cluster ($H(\boldsymbol{\theta}_j, \boldsymbol{\theta}_i)$ achieves the lowest distance, when $\boldsymbol{\theta}_j = \boldsymbol{\theta}_i$ and $\theta_{i,k} = 1$ for some cluster $k$), which agrees with our clustering purpose, we pick it over KL-divergence.

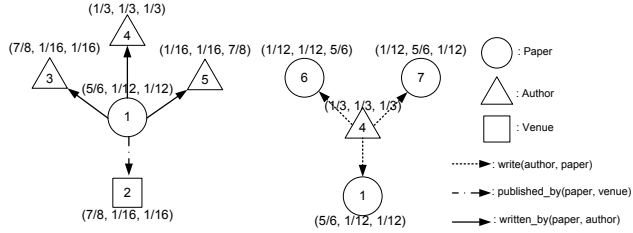

**Figure 4: Illustration of Feature Function**

Fig. 4 illustrates a small example of a bibliographic network containing 7 objects. For clarity, we only draw the out-links of two objects corresponding to Paper 1 and Author 4. The weights of all links are 1, and the given membership vector with respect to three clusters is shown in the figure. Three link types are contained in the network, corresponding to $write(author, paper)$ with strength weight $\gamma_1$, $published\_by(paper, venue)$ with weight $\gamma_2$, and $written\_by(paper, author)$ with weight $\gamma_3$. From the example, we can see that:

1. Objects 1 and 3 are more likely to belong to the first cluster, Object 4 is a neutral object, and Object 5 is more likely to belong to the third cluster. With Eq. (6), we get $f(\langle 1, 3 \rangle) = -0.4701\gamma_3$; $f(\langle 1, 4 \rangle) = -1.7174\gamma_3$; and $f(\langle 1, 5 \rangle) = -2.3410\gamma_3$. In other words, $f(\langle 1, 3 \rangle) \geq f(\langle 1, 4 \rangle) \geq f(\langle 1, 5 \rangle)$. This satisfies the first desired criterion.

2. $f(\langle 1, 2 \rangle) = -0.4701\gamma_2$ and $f(\langle 1, 3 \rangle) = -0.4701\gamma_3$. If $\gamma_2 < \gamma_3$, (or, the strength of link type $published\_by$ is smaller than $written\_by$), then $f(\langle 1, 2 \rangle) > f(\langle 1, 3 \rangle)$. That is to say, in order to obtain the same value for two feature functions defined on two different link types, the link type with stronger strength requires even greater similarity for the membership vectors. In other words, stronger link types are likely to exist only between objects that are very similar to each other, and indicate a better quality of the link type.

3. $f(\langle 1, 4 \rangle) = -1.7174\gamma_3$, $f(\langle 4, 1 \rangle) = -1.0986\gamma_1$, and in general $f(\langle 1, 4 \rangle) \neq f(\langle 4, 1 \rangle)$. Even if the two links belong to the same type, i.e., $\gamma_3 == \gamma_1$, we still have $f(\langle 1, 4 \rangle) < f(\langle 4, 1 \rangle)$. The intuitive explanation is that it is less helpful for a neutral object to decide an object's expertise than for an expert object to decide whether an object is neutral. Therefore, the asymmetric criterion holds as well.

We then propose a log-linear model to model the probability of $\Theta$ given the link type weights $\boldsymbol{\gamma}$, where the probability of one configuration $\Theta$ is defined as the exponential of the summation of feature functions of all the links in $G$:

$$p(\Theta | G, \boldsymbol{\gamma}) = \frac{1}{Z(\boldsymbol{\gamma})} \exp\{ \sum_{e=\langle v_i, v_j \rangle \in E} f(\boldsymbol{\theta}_i, \boldsymbol{\theta}_j, e, \boldsymbol{\gamma}) \} \quad (7)$$

where $\boldsymbol{\gamma}$ is the strength weight vector for all link types, $f(\boldsymbol{\theta}_i, \boldsymbol{\theta}_j, e, \boldsymbol{\gamma})$ is the feature function defined on links of different types, and $Z(\boldsymbol{\gamma})$ is the partition function that makes the distribution function sum up to 1: $Z(\boldsymbol{\gamma}) = \int_{\Theta} \exp\{\sum_{e=\langle v_i, v_j \rangle \in E} f(\boldsymbol{\theta}_i, \boldsymbol{\theta}_j, e, \boldsymbol{\gamma})\} d\Theta$. The partition function $Z(\boldsymbol{\gamma})$ is an integral over the space of all the configurations $\boldsymbol{\Theta}$, and it is a function of $\boldsymbol{\gamma}$.

### 3.4 The Unified Model

The overall goal of the network clustering problem is to determine the best clustering results $\Theta$, the link type strengths $\boldsymbol{\gamma}$ and the cluster component parameters $\boldsymbol{\beta}$ that maximize the generative probability of attribute observations and the consistency with the network structure, described by the likelihood function in Eq. (1).

Further, we add a Gaussian prior to $\boldsymbol{\gamma}$ as a regularization to avoid overfitting, with the mean as 0, and the covariance matrix as $\sigma^2 I$, where $\sigma$ is the standard deviation of each element in $\boldsymbol{\gamma}$, and $I$ is the identity matrix. We set $\sigma = 0.1$ in our experiments, and more complex strategy can be used to select $\sigma$ according to labeled clustering results, which will not be discussed here. The new objective function is then:

$$g(\Theta, \boldsymbol{\beta}, \boldsymbol{\gamma}) = \log \sum_{X \in \mathcal{X}} p(\{v[X]\}_{v \in V_X} | \Theta, \boldsymbol{\beta}) + \log p(\Theta | G, \boldsymbol{\gamma}) - \frac{||\boldsymbol{\gamma}||^2}{2\sigma^2} \quad (8)$$

In addition, we have the constraints that $\boldsymbol{\gamma} \geq 0$, and some constraints for $\boldsymbol{\beta}$ that are dependent on the attribute distribution type. Also, $p(\{v[X]\}_{v \in V_X} | \Theta, \boldsymbol{\beta})$ and $p(\Theta | G, \boldsymbol{\gamma})$ need to be replaced by the specific formulas proposed above for concrete derivations.

## 4. THE CLUSTERING ALGORITHM

This section presents a clustering algorithm that computes the proposed probabilistic clustering model. Intuitively, we begin with the assumption that all the types of links play an equally important role in the clustering process, then update the strength for each type according to the average consistency of links of that type with the current clustering results, and finally achieve a good clustering as well as a reasonable strength vector for link types. It is an iterative algorithm containing two steps in that clustering results and strengths of link types mutually enhance each other, which maximizes the objective function of Eq. (8) alternatively.

In the first step, we fix the link type weights $\boldsymbol{\gamma}$ to the best value $\boldsymbol{\gamma}^*$, determined in the last iteration, then the problem becomes that of determining the best clustering results $\Theta$ and the attribute parameters $\beta$ for each cluster component. We refer to this step as the *cluster optimization step*: $[\Theta^*, \beta^*] = \arg\max_{\Theta, \beta} g(\Theta, \beta, \gamma^*)$.

In the second step, we fix the clustering configuration parameters $\Theta = \Theta^*$ and $\beta = \beta^*$, corresponding to the values determined in the last step, and use it to determine the best value of $\gamma$, which is consistent with current clustering results. We refer to this step as the *link type strength learning step*: $\gamma^* = \arg\max_{\gamma \geq 0} g(\Theta^*, \beta^*, \gamma)$.

The two steps are repeated until convergence is achieved.

## 4.1 Cluster Optimization

In the cluster optimization step, each object has the link information from different types of neighbors, where the strength of each type of link is given, as well as the possible attribute observations. The goal is to utilize both link and attribute information to get the best clustering for all the objects. Since $\gamma$ is fixed in this step, the partition function and regularizer term become constants, and can be discarded for optimization purposes. Therefore, we can construct a simplified objective function $g_1(\cdot, \cdot)$, which depends only on $\Theta$ and $\beta$:

$$g_1(\Theta, \beta) = \sum_{e=\langle v_i, v_j \rangle} f(\theta_i, \theta_j, e, \gamma) + \sum_{v \in V_X} \sum_{x \in v[X]} \log \sum_{k=1}^{K} \theta_{v,k} p(x|\beta_k)$$
(9)

We derived an EM-based algorithm [9, 4] to solve Eq. (9). In the E-step, the probability of each observation $x$ for each object $v$ and each attribute $X$ belonging to each cluster, usually called the hidden cluster label of the observation, $z_{v,x}$, is derived according to the current parameters $\Theta$ and $\beta$. In the M-step, the parameters $\Theta$ and $\beta$ are updated according to the new membership for all the observations in the E-step. The iterative formulas for single text attribute, single Gaussian attribute, and two Gaussian attributes are provided below.

**1. Single Categorical text attribute:** Let $z_{v,l}$ denote the hidden cluster label for the $l$th term in the vocabulary for object $v$, $\Theta^{t-1}$ be the value of $\Theta$ at iteration $t-1$, and $\beta^{t-1}$ be the value of $\beta$ at iteration $t-1$. $\mathbf{1}_{\{v \in V_X\}}$ is the indicator function, which is 1 if $v$ contains this attribute, otherwise 0. Then, we have:

$$p(z_{v,l}^t = k | \Theta^{t-1}, \beta^{t-1}) \propto \theta_{v,k}^{t-1} \beta_{k,l}^{t-1}$$

$$\theta_{v,k}^t \propto \sum_{e=\langle v,u \rangle} \gamma(\phi(e)) w(e) \theta_{u,k}^{t-1} + \mathbf{1}_{\{v \in V_X\}} \sum_{l=1}^{m} c_{v,l} p(z_{v,l}^t = k | \Theta^{t-1}, \beta^{t-1})$$

$$\beta_{k,l}^t \propto \sum_{v \in V_X} c_{v,l} p(z_{v,l}^t = k | \Theta^{t-1}, \beta^{t-1})$$
(10)

**2. Single Gaussian numerical attribute:** Let $z_{v,x}$ denote the hidden cluster label for the observation $x$ for object $v$, $\Theta^t$ be the value of $\Theta$ at iteration $t$, and $\mu_k^t$ and $\sigma_k^t$ be the values of mean and standard deviation for $k$th cluster at iteration $t$. $\mathbf{1}_{\{v \in V_X\}}$ is the indicator function, which is 1 if $v$ contains this attribute, otherwise 0. Then, we have:

$$p(z_{v,x}^t = k | \Theta^{t-1}, \beta^{t-1}) \propto \theta_{v,k}^{t-1} \frac{1}{\sqrt{2\pi(\sigma_k^{t-1})^2}} e^{-\frac{(x-\mu_k^{t-1})^2}{2(\sigma_k^{t-1})^2}}$$

$$\theta_{v,k}^t \propto \sum_{e=\langle v,u \rangle} \gamma(\phi(e)) w(e) \theta_{u,k}^{t-1} + \mathbf{1}_{\{v \in V_X\}} \sum_{x \in v[X]} p(z_{v,x}^t = k | \Theta^{t-1}, \beta^{t-1})$$

$$\mu_k^t = \frac{\sum_{v \in V_X} \sum_{x \in v[X]} x p(z_{v,x}^t = k | \Theta^{t-1}, \beta^{t-1})}{\sum_{v \in V_X} \sum_{x \in v[X]} p(z_{v,x}^t = k | \Theta^{t-1}, \beta^{t-1})}$$

$$(\sigma_k^2)^t = \frac{\sum_{v \in V_X} \sum_{x \in v[X]} (x - \mu_k^t)^2 p(z_{v,x}^t = k | \Theta^{t-1}, \beta^{t-1})}{\sum_{v \in V_X} \sum_{x \in v[X]} p(z_{v,x}^t = k | \Theta^{t-1}, \beta^{t-1})}$$
(11)

**3. Two Gaussian numerical attributes:** Let $X, Y$ be two attributes following Gaussian distributions, $z_{v,x}, z_{v,y}$ denote the hidden cluster labels of the observation $x$ for attribute $X$ and the observation $y$ for attribute $Y$ respectively for object $v$, $\Theta^t$ be the value of $\Theta$ at iteration $t$, and $\mu_{X,k}^t, \mu_{Y,k}^t$ and $\sigma_{X,k}^t, \sigma_{Y,k}^t$ be the values of mean and standard deviation for $k$th cluster of attribute $X$ and $Y$ at iteration $t$. $\mathbf{1}_{\{v \in V_X\}}$ and $\mathbf{1}_{\{v \in V_Y\}}$ are the indicator functions, which are 1 if $v$ contains $X$ or $Y$, otherwise 0. Then, we have:

$$p(z_{v,x}^t = k | \Theta^{t-1}, \beta^{t-1}) \propto \theta_{v,k}^{t-1} \frac{1}{\sqrt{2\pi(\sigma_{X,k}^{t-1})^2}} e^{-\frac{(x-\mu_{X,k}^{t-1})^2}{2(\sigma_{X,k}^{t-1})^2}}$$

$$p(z_{v,y}^t = k | \Theta^{t-1}, \beta^{t-1}) \propto \theta_{v,k}^{t-1} \frac{1}{\sqrt{2\pi(\sigma_{Y,k}^{t-1})^2}} e^{-\frac{(y-\mu_{Y,k}^{t-1})^2}{2(\sigma_{Y,k}^{t-1})^2}}$$

$$\theta_{v,k}^t \propto \sum_{e=\langle v,u \rangle} \gamma(\phi(e)) w(e) \theta_{u,k}^{t-1} + \mathbf{1}_{\{v \in V_X\}} \sum_{x \in v[X]} p(z_{v,x}^t = k | \Theta^{t-1}, \beta^{t-1}) + \mathbf{1}_{\{v \in V_Y\}} \sum_{y \in v[Y]} p(z_{v,y}^t = k | \Theta^{t-1}, \beta^{t-1})$$

$$\mu_{X,k}^t = \frac{\sum_{v \in V_X} \sum_{x \in v[X]} x p(z_{v,x}^t = k | \Theta^{t-1}, \beta^{t-1})}{\sum_{v \in V_X} \sum_{x \in v[X]} p(z_{v,x}^t = k | \Theta^{t-1}, \beta^{t-1})}$$

$$(\sigma_{X,k}^2)^t = \frac{\sum_{v \in V_X} \sum_{x \in v[X]} (x - \mu_{X,k}^t)^2 p(z_{v,x}^t = k | \Theta^{t-1}, \beta^{t-1})}{\sum_{v \in V_X} \sum_{x \in v[X]} p(z_{v,x}^t = k | \Theta^{t-1}, \beta^{t-1})}$$

$$\mu_{Y,k}^t = \frac{\sum_{v \in V_Y} \sum_{y \in v[Y]} y p(z_{v,y}^t = k | \Theta^{t-1}, \beta^{t-1})}{\sum_{v \in V_Y} \sum_{y \in v[Y]} p(z_{v,y}^t = k | \Theta^{t-1}, \beta^{t-1})}$$

$$(\sigma_{Y,k}^2)^t = \frac{\sum_{v \in V_Y} \sum_{y \in v[Y]} (y - \mu_{Y,k}^t)^2 p(z_{v,y}^t = k | \Theta^{t-1}, \beta^{t-1})}{\sum_{v \in V_Y} \sum_{y \in v[Y]} p(z_{v,y}^t = k | \Theta^{t-1}, \beta^{t-1})}$$
(12)

A more detailed derivation of the EM algorithm is provided for single text attribute in Appendix A, which is similar for single or multiple Gaussian numerical attributes.

From the update rules, we can see that the value of the membership probability for an object is dependent on its neighbors' memberships, the strength of the link types, the weight of the links, and the attribute associated with it (if any). When an object contains no attributes in the specified set, or contains no observations for the specified attributes, the cluster membership is totally determined by its linked objects, which is a weighted average of their cluster memberships and the weight is determined by both the weight of the link and the weight of the link type. When an object contains some observations of the specified attributes, its cluster membership is determined by both its neighbors and these observations for each possible attribute.

## 4.2 Link Type Strength Learning

The link type strength learning step is to find the best strength weight for each type of links that makes the current clustering result to be generated with the highest probability. By doing so, the low quality link types that connect objects not so similar will be punished and assigned with low strength weights; while the high quality link types will be assigned with high strength weights.

Since the values of $\Theta$ and $\beta$ are fixed in this step, the only relevant parts of the objective function (for optimization purposes) are those which depend on $\gamma$. These are the structural consistency modeling part and the regularizer over $\gamma$. Therefore, we can construct the following simplified objective function $g_2(\cdot)$ as a function of $\gamma$:

$$g_2(\gamma) = \sum_{e=\langle v_i, v_j \rangle} f(\theta_i, \theta_j, e, \gamma) - \log Z(\gamma) - \frac{||\gamma||^2}{2\sigma^2} \quad (13)$$

In addition, we have the linear constraints as $\gamma \geq 0$.

However, $g_2$ is difficult to be optimized directly, since the partition function $Z(\gamma)$ is an integral over the entire space of valid values of $\Theta$, which is intractable. Instead, we construct an alternate approximate objective function $g_2'$, which factorizes $\log p(\Theta|G)$ as the sum of $\log p(\theta_i|\theta_{-i}, G)$, namely the pseudo-log-likelihood, where $p(\theta_i|\theta_{-i}, G)$ is the conditional probability of $\theta_i$ given the remaining objects' clustering configurations, which turns out to be dependent only on its neighbors. The intuition of using pseudo-log-likelihood to approximate the real log-likelihood is that, if the probability of generating the clustering configuration for each object conditional on its neighbors is high, the probability of generating the whole clustering configuration should also be high. In other words, if the local patches of a network are very consistent with the clustering results, the consistency over the whole network should also be high.

In particular, we choose each local patch of the network as an object and all its out-link neighbors. In this case, every link is considered exactly once, and the newly designed objective function $g_2'(\cdot)$ is as follows:

$$g_2'(\gamma) = \sum_{i=1}^{|V|} \left( \sum_{e=\langle v_i, v_j \rangle} f(\theta_i, \theta_j, e, \gamma) - \log Z_i(\gamma) \right) - \frac{||\gamma||^2}{2\sigma^2} \quad (14)$$

where $\log Z_i(\gamma) = \log \int_{\theta_i} e^{\sum_{e=\langle v_i, v_j \rangle} f(\theta_i, \theta_j, e, \gamma)} d\theta_i$, the local partition function for object $v_i$, with the linear constraints $\gamma \geq 0$.

As the joint distribution of $\Theta$ as well as the conditional distribution of $\theta_i$ given its out-link neighbors are both belonging to exponential families, both $g_2$ and $g_2'$ are concave functions of $\gamma$, and the concavity of $g_2'$ is proved in Appendix B. Therefore, the maximum value is either achieved at the global maximum point or at the boundary of constraints. The Newton-Raphson method is used to solve the optimization problem. It needs to calculate the first and second derivative of $g_2'(\gamma)$ with respect to $\gamma$, which is non-trivial in our case. We discuss the computation of these below.

By re-examining $p(\theta_i|\{\theta_j\}_{\forall e=\langle v_i, v_j \rangle}, G)$, the conditional probability for each object $i$ given its out-link neighbors, we have:

$$p(\theta_i|\{\theta_j\}_{\forall e=\langle v_i, v_j \rangle}, G) \propto \prod_{k=1}^{K} \theta_{ik}^{\sum_{e=\langle v_i, v_j \rangle} \gamma(\phi(e))w(e)\theta_{j,k}} \quad (15)$$

It is easy to see that $p(\theta_i|\{\theta_j\}_{\forall e=\langle v_i, v_j \rangle}, G)$ is a Dirichlet distribution with parameters $\alpha_{ik} = \sum_{e=\langle v_i, v_j \rangle} \gamma(\phi(e))w(e)\theta_{j,k} + 1$, for $k = 1$ to $K$. Therefore, the local partition function for each object $i$, $Z_i(\gamma)$, should be the constant $B(\alpha_i)$ as in Dirichlet distribution, where $\alpha_i = (\alpha_{i1}, \ldots, \alpha_{iK})$ and $B(\alpha_i) = \frac{\prod_{k=1}^{K} \Gamma(\alpha_{ik})}{\Gamma(\sum_{k=1}^{K} \alpha_{ik})}$. Then the first and second derivatives ($\nabla g_2'$ and $Hg_2'$) can be calculated now as each $Z_i$ is a function of Gamma functions.

The first derivative (or gradient) of $g_2'$ is expressed as:

$$\nabla g_2'(r) = \sum_{i=1}^{|V|} \left( \sum_{\substack{e=\langle v_i, v_j \rangle \\ \phi(e)=r}} w(e) \sum_{k=1}^{K} \theta_{jk} \log \theta_{ik} \right.$$
$$\left. - \left( \sum_{k=1}^{K} \psi(\alpha_{ik}) \sum_{\substack{e=\langle v_i, v_j \rangle \\ \phi(e)=r}} w(e)\theta_{jk} - \psi(\sum_{k=1}^{K} \alpha_{ik}) \sum_{\substack{e=\langle v_i, v_j \rangle \\ \phi(e)=r}} w(e) \right) \right) - \frac{\gamma(r)}{\sigma^2} \quad (16)$$

for every $r \in \mathcal{R}$, where $\psi(x)$ is the digamma function that is the first derivative of $\log \Gamma(x)$, namely $\psi(x) = \Gamma'(x)/\Gamma(x)$.

---

**Input**: Network $G$, Attribute $X_1, \ldots, X_T$, cluster number K;
**Output**: Cluster membership $\Theta$, Link type weights $\gamma$, attribute component parameters $\beta_1, \ldots, \beta_T$;

Initialization for $\gamma^0$;
**repeat**
  %Step 1: Optimization of $\Theta^t$ given $\gamma^{t-1}$;
  Initialize $\Theta'^0, \beta'^0$;
  **repeat**
    1. for each object $v$, update $p(z_{v,x}^s = k|\Theta'^{s-1}, \beta'^{s-1})$;
    2. for each object $v$, update $\theta_{v,k}^s$;
    3. for each cluster $k$, update parameter for each attribute $X_i, \beta'_{i,k}$;
  **until** *reaches precision requirement for $\Theta'^s$*;
  $\Theta^t = \Theta'^s$;
  $\beta^t = \beta'^s$;
  %Step 2: Optimization of $\gamma^t$ given $\Theta^t$;
  $\gamma'^0 = \gamma^{t-1}$;
  **repeat**
    1. $\gamma'^s = \gamma'^{s-1} - [Hg_2'(\gamma'^{s-1})]^{-1} \nabla g_2'(\gamma'^{s-1})$;
    2. $\forall r \in \mathcal{R}$, if $\gamma'(r)^s < 0$, set $\gamma'(r)^s = 0$;
  **until** *reaches precision requirement for $\gamma'^s$*;
  $\gamma^t = \gamma'^s$;
**until** *reaches iteration number or precision requirement for $\gamma^t$*;

**Algorithm 1**: The *GenClus* Algorithm.

The second derivative (or Hessian matrix) of $g_2'$, can be expressed as:

$$Hg_2'(r_1, r_2) = \sum_{i=1}^{n} \left( -\sum_{k=1}^{K} \psi'(\alpha_{ik}) \sum_{\substack{e=\langle v_i, v_j \rangle \\ \psi(e)=r_1}} w(e)\theta_{jk} \sum_{\substack{e=\langle v_i, v_j \rangle \\ \psi(e)=r_2}} w(e)\theta_{jk} \right.$$
$$\left. + \psi'(\sum_{k=1}^{K} \alpha_{ik}) \sum_{\substack{e=\langle v_i, v_j \rangle \\ \psi(e)=r_1}} w(e) \sum_{\substack{e=\langle v_i, v_j \rangle \\ \psi(e)=r_2}} w(e) \right) - \frac{1}{\sigma^2} \mathbf{1}_{\{r_1=r_2\}} \quad (17)$$

for every pair of relations $r_1, r_2 \in \mathcal{R}$, where $\psi'(x)$ is the first derivative of $\psi(x)$, and $\mathbf{1}_{\{r_1=r_2\}}$ is the indicator function, with the value 1 if $r_1 = r_2$, and 0 otherwise.

Then, we can use the Newton-Raphson method to determine the value of $\gamma$ that maximizes $g_2'$ with the following iterative steps:

1. $\gamma^{t+1} = \gamma^t - [Hg_2'(\gamma^t)]^{-1} \nabla g_2'(\gamma^t)$;

2. $\forall r \in \mathcal{R}$, if $\gamma(r)^{t+1} < 0$, set $\gamma(r)^{t+1} = 0$.

### 4.3 Putting together: The GenClus Algorithm

We integrate the two steps discussed above to construct a **Gen**eral Heterogeneous Network **Clus**tering algorithm, *GenClus*, as shown in Algorithm 1 in pseudo code.

The algorithm includes an outer iteration that updates $\Theta$ and $\gamma$ alternatively, and two inner iterations that optimize $\Theta$ using the EM algorithm and optimize $\gamma$ using the Newton-Raphson method respectively. For the initialization of $\gamma$ in the outer iteration, we initialize it as an all-1 vector. This means that all the link types in the network are initially considered equally important. For the initialization of $\Theta'$ in the inner iteration for optimizing $\Theta$, we can either (1) assign $\Theta'^0$ with random assignments, or (2) start with several random seeds, run the EM algorithm for a few steps for each random seed, and choose the one with the highest value of the objective function $g_1$ as the real starting point. The latter approach will produce more stable results.

The time complexity for the EM algorithm in the first step is $O(t_1(Kd_1|V| + K|E|))$, where $t_1$ is the number of iterations, $d_1$ is the average number of observations for each object, $K$ is the number of clusters, $|V|$ is the number of objects, and $|E|$ is the number

400

of links in the network, which is linear to $|V|$ for sparse networks. The time complexity of the algorithm in the step of maximizing $\gamma$ is dependent on the time for calculating the first derivative and Hessian matrix of $g_2'(\gamma)$, and the matrix inversion involved Newton-Raphson algorithm. This is $O(K|E| + t_2|R|^{2.376}))$, where $K$ and $|E|$ are with the same meaning as before, $t_2$ is the number of iterations, and $|R|$ is the number of relations in the network. In all, the overall time complexity is $O(t(t_1(Kd_1|V|+K|E|)+t_2|R|^{2.376}))$, where $t$ is the number of outer iterations. In other words, for each outer iteration, the time complexity is approximately linear in the number of objects in the network when the network is sparse. Therefore, the *GenClus* algorithm is quite scalable.

## 5. EXPERIMENTAL RESULTS

In this section, we examine the effectiveness and efficiency of the clustering algorithm on several real and synthetic data sets.

## 5.1 Data Sets

Two real networks and one synthetic network are used in this study. From the *DBLP Four-area data set* [23] [10], we extracted two networks where the network structures are represented by different subsets of entities and their corresponding links. This data set was extracted from 20 major conferences from the four areas corresponding to database, data mining, machine learning, and information retrieval. It contains 14376 papers and 14475 authors, corresponding to publications before 2008. Labels were associated with a subset of the nodes, and specifically with 20 conferences, 100 papers, and 4236 authors into 4 areas. Besides the real networks, we also generated a synthetic weather sensor network. We describe these networks below:

**(a) DBLP Four-area AC Network.** This network contains two types of objects, authors (A) and conferences (C); and three types of links depending upon publication behavior, namely $publish\_in(A,C)$ (abbr. as $\langle A,C \rangle$), $published\_by(C,A)$ (abbr. as $\langle C,A \rangle$), and $coauthor(A,A)$ (abbr. as $\langle A,A \rangle$). The links are associated with a weight corresponding to the number of papers that an author has published in a conference, a conference is contributed by an author, and the two authors have coauthored, respectively. The author nodes and conference nodes contain text corresponding to the text from the titles of all the papers they have ever written or published.

**(b) DBLP Four-area ACP Network.** This network contains objects corresponding to authors (A), conferences (C) and papers (P); and four types of links depending upon the publication behavior, namely $write(A,P)$ (abbr. as $\langle A,P \rangle$), $written\_by(P,A)$ (abbr. as $\langle P,A \rangle$), $publish(C,P)$ (abbr. as $\langle C,P \rangle$), and $published\_by(P,C)$ (abbr. as $\langle P,C \rangle$). In this case, the links have binary weights, corresponding to presence or absence of the link. Only papers contain text attributes, extracted from their titles.

**(c) Weather Sensor Network.** This network is synthetically generated, containing two types of objects: temperature (T) and precipitation (P) sensors, and four link types between any two types of sensors denoting the kNN relationship: $\langle T,T \rangle, \langle T,P \rangle, \langle P,T \rangle$, and $\langle P,P \rangle$. The links are binary weighted according to their $k$-nearest neighbors. The attributes associated with a sensor correspond to either temperature or precipitation, depending on the type of the sensor.

The weather sensor network is generated by assuming there are $K$ weather patterns, each of which is defined as a Gaussian distribution over temperature and precipitation attributes with different parameters. The links are built according to the $k$-nearest neighbors relationship. The temperature and precipitation observations are generated by sampling. The details of the sensor network generator is introduced in Appendix C. We use the weather network generator to generate two sets of synthetic climate sensor networks, each containing 4 clusters, and each sensor is linked to 5 nearest neighbors for each type (10 in total). The first set of networks have attribute means as $(1,1), (2,2), (3,3), (4,4)$ for each cluster, and the standard deviation for both attributes is set to 0.2. The correlation between temperature and precipitation is 0. The second set of networks have attribute means as $(1,1), (-1,1), (-1,-1), (1,-1)$ for each cluster, with the same covariance matrix as the first setting. Notice that Setting 2 is more difficult than Setting 1, in the sense that the weather pattern can only be determined when we know both the temperature and precipitation observations for each location. The temperature sensors have soft cluster membership in two neighboring clusters (less noisy); while precipitation sensors have soft membership in three neighboring clusters (more noisy). In each setting, we vary the number of sensors, by fixing the number of temperature sensors as 1000, and precipitation sensors as 250, 500, and 1000. For each setting, the number of observations for each object may be 1, 5 or 20. In all, for each weather pattern setting, we have 9 networks with different configurations.

## 5.2 Effectiveness Study

We use two measures for our effectiveness study. First, the labels associated with the nodes in the data sets provide a natural guidance in examining the coherence of the clusters. We use *Normalized Mutual Information (NMI)* [21] to compare our clustering result with the ground truth, which evaluates the similarity between two partitions of the objects. Second, we use link prediction accuracy to test the clustering accuracy. The similarity between two objects can be calculated by a similarity function defined on their two membership vectors, such as cosine similarity. Clearly, a better clustering quality leads to better computation of similarity (and therefore the better accuracy of link prediction). For a certain type of relation $\langle A,B \rangle$, we calculate the similarity scores between each $v_A \in A$ and all the objects $v_B \in B$, and compare the similarity-based ranked list with the true ranked list determined by the link weights between them. We use the measure *Mean Average Precision (MAP)* [27] to compare the two ranked links.

### 5.2.1 Clustering Accuracy Test

We choose clustering methods that can deal with both links and attributes as our baselines. None of these baselines is capable of leveraging different link types in terms of their differential impact to the clustering process. Therefore, we set each link type strength as 1 for these baselines. Second, we choose different baselines for clustering networks with text attributes and clustering networks with numerical attributes, since there is no unified clustering method (other than our presented *GenClus*) that can address both situations in the same framework.

For *DBLP Four-area AC Network* and *DBLP Four-area ACP Network* that are with text attributes, we use *NetPLSA* [18] and *iTopicModel* [22] as baselines, which aim at improving topic qualities by using link information in homogeneous networks. We compare *GenClus* with these baselines by assuming homogeneity of links for the latter two algorithms. The number of iterations of GenClus is set to 10. Each algorithm is run for 20 times with random initial settings. The mean and standard deviation of NMI of the 20 running results are shown for the *DBLP AC Network* and *DBLP ACP Network* in Figs. 5 and 6 respectively. From the results, we can see that *GenClus* is much more effective than *iTopicModel* and *NetPLSA* in both networks, due to the ability of *GenClus* to learn and leverage the strengths of different link types in the clustering process. Furthermore, the standard deviation of NMI over differ-ent runs is much lower for *GenClus*, which suggests that the algorithm is more robust to the initial settings with the learned strength weights for different link types.

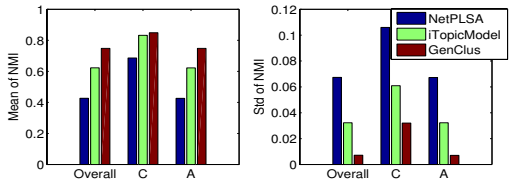

**Figure 5: Clustering Accuracy Comparisons for AC Network**

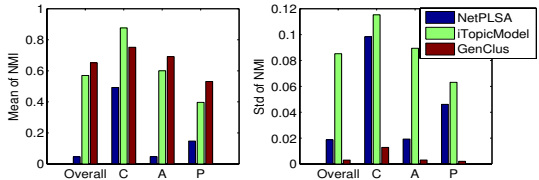

**Figure 6: Clustering Accuracy Comparisons for ACP Network**

The *AC Network* is the easiest case among the three networks, since it only contains one type of attribute (the text attribute), and all object types contain this attribute, namely, the attribute is complete for every object. The *ACP network* is a more difficult case, because not every type of objects contains the text attributes. This requires the clustering algorithm to be more robust to deal with objects with no attributes. From the results, we can see that *GenClus* is more robust than *NetPLSA*, which outputs almost random predictions for authors for the ACP network. Although *iTopicModel* performs better for objects of type C for the ACP network (see Fig. 6), *GenClus* still has an overall better performance. This is because our objective function is defined over all the object types rather than on a particular type.

We also examined the actual clusters obtained by the algorithm on the DBLP AC network, and list the corresponding cluster memberships for several well-known conferences and authors in Table 1, where the research area names are given afterwards according to the clustering results. We can see that the clustering results of *GenClus* are consistent with human intuition.

| Object | DB | DM | IR | ML |
|---|---|---|---|---|
| SIGMOD | 0.8577 | 0.0492 | 0.0482 | 0.0449 |
| KDD | 0.0786 | 0.6976 | 0.1212 | 0.1026 |
| CIKM | 0.2831 | 0.1370 | 0.4827 | 0.0971 |
| Jennifer Widom | 0.7396 | 0.0830 | 0.1061 | 0.0713 |
| Jim Gray | 0.8359 | 0.0656 | 0.0536 | 0.0449 |
| Christos Faloutsos | 0.4268 | 0.3055 | 0.1380 | 0.1296 |

**Table 1: Case Studies of Cluster Membership Results**

The synthetic weather sensor network is the most difficult case among the three networks, as it has two types of attributes corresponding to different types of sensors. Furthermore, all sensor nodes contain incomplete attributes. Existing algorithms cannot address these issues well. We compare the clustering results of *GenClus* with two baselines, by comparing the cluster labels with maximum probabilities with the ground truth. In this case, we choose the initial seed for *GenClus* as one of the tentative running results with the highest objective function, and the iteration number is set to 5. The first baseline is the $k$-means algorithm, and the second is a spectral clustering method that combines the network structure and attribute similarity as a new similarity matrix. We use the framework given in [20], which utilizes modularity objective function in the network part, but we replace the cosine similarity by Euclidean distance in the attribute part as in [26] for better clustering results. As neither methods can handle the problem of incomplete attributes, we use interpolation to make each sensor have a regular 2-dimensional attribute, by using the mean of all the observations of its neighbors and itself. For the spectral clustering-based framework, we centralize the data by extracting the mean and then normalize them by the standard deviation, in order to make the attribute part comparable with the modularity part in the objective function. Both parts are set to have equal weights.

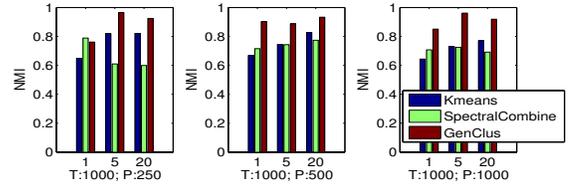

**Figure 7: Clustering Accuracy Comparisons for Setting 1**

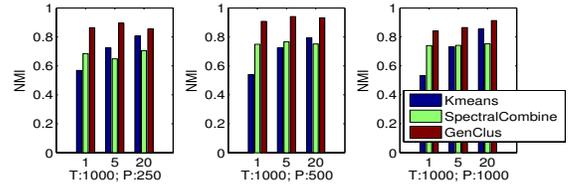

**Figure 8: Clustering Accuracy Comparisons for Setting 2**

The results are summarized in Figs. 7 and 8. It is evident that *GenClus* exhibits superior performance over the two baselines in most of the data sets (17 out of 18 cases). Furthermore, *GenClus* can produce more stable clustering results compared with $k$-means, which is very sensitive to the number of observations for each object, especially for Setting 2. *GenClus* is also highly adaptive: no need of any weight specification for combining the network and attribute-contributions to the clustering process. This results in greater stability for the *GenClus* algorithm. Another major advantage of *GenClus* (which is not immediately evident from the presented results) is that we can directly utilize every observation instead of the mean, whereas the baselines can only use a biased mean value because of the interpolation process.

### 5.2.2 Link Prediction Accuracy Test

Next, the link prediction accuracy measured by MAP is compared between *GenClus* and the baselines. For the AC network, we select the link type $\langle A, C \rangle$ for prediction, namely, we want to predict which conferences that an author is likely to publish in. For the APC network, we select the link type $\langle P, C \rangle$ for prediction, namely, we want to predict which conference that a paper is published in. As the prediction is based on the similarity between the two objects, say query object $v_i$ with clustering membership $\boldsymbol{\theta}_i$ and candidate object $v_j$ with clustering membership $\boldsymbol{\theta}_j$, three similarity functions are used here: (1) cosine similarity denoted as $\cos(\boldsymbol{\theta}_i, \boldsymbol{\theta}_j)$; (2) the negative of Euclidean distance denoted as $-||\boldsymbol{\theta}_i - \boldsymbol{\theta}_j||$; and (3) the negative of cross entropy denoted as $-H(\boldsymbol{\theta}_j, \boldsymbol{\theta}_i)$. The results are summarized in Tables 2 and 3.

|   | NetPLSA | iTopicModel | GenClus |
|---|---|---|---|
| $\cos(\boldsymbol{\theta}_i, \boldsymbol{\theta}_j)$ | 0.4351 | 0.5117 | **0.7627** |
| $-\|\boldsymbol{\theta}_i - \boldsymbol{\theta}_j\|$ | 0.4312 | 0.5010 | **0.7539** |
| $-H(\boldsymbol{\theta}_j, \boldsymbol{\theta}_i)$ | 0.4323 | 0.5088 | **0.7753** |

**Table 2: Prediction Accuracy for A-C Relation in AC Network**

|   | NetPLSA | iTopicModel | GenClus |
|---|---|---|---|
| $\cos(\boldsymbol{\theta}_i, \boldsymbol{\theta}_j)$ | 0.2762 | 0.4609 | **0.5170** |
| $-\|\boldsymbol{\theta}_i - \boldsymbol{\theta}_j\|$ | 0.2759 | 0.4600 | **0.5142** |
| $-H(\boldsymbol{\theta}_j, \boldsymbol{\theta}_i)$ | 0.2760 | 0.4683 | **0.5183** |

**Table 3: Prediction Accuracy for P-C Relation in ACP Network**

For the weather sensor network, we select the link type $\langle T, P \rangle$, namely, we want to predict the P-typed neighbors for the T-typed sensors. We test the link prediction in the network with configuration as in Setting 1, with $\#T = 1000$ and $\#P = 250$. We only output the link prediction results for the *GenClus* algorithm, since the other two baselines can only output hard clusters (exact cluster memberships rather than probabilities). The results are shown in Table 4.

|   | $\cos(\boldsymbol{\theta}_i, \boldsymbol{\theta}_j)$ | $-\|\boldsymbol{\theta}_i - \boldsymbol{\theta}_j\|$ | $-H(\boldsymbol{\theta}_j, \boldsymbol{\theta}_i)$ |
|---|---|---|---|
| MAP | 0.7285 | 0.7690 | **0.8073** |

**Table 4: Prediction Accuracy for $\langle T, P \rangle$ in Weather Network**

From the results, it is evident that *GenClus* has the best link prediction accuracy in terms of different similarity functions. Also, the results show that the asymmetric function $-H(\boldsymbol{\theta}_j, \boldsymbol{\theta}_i)$ provides the best link prediction accuracy, especially for better clustering results such as those obtained by *GenClus* and in the weather sensor network where the out-link neighbors are different from the in-link neighbors.

### 5.2.3 *Analysis of Link Type Strength*

Since the process of learning the semantic importance of relations is important in a heterogeneous clustering approach, we present the learned relation strengths in Fig. 9 for the two DBLP four-area networks. From the figure, it is evident that in the AC Network, the link type $\langle A, C \rangle$ has greater importance to the clustering process than the link type $\langle A, A \rangle$, and thus is more important in deciding an author's membership. This is because the spectrum of co-authors may often be quite broad, whereas their publication frequency in each conference can be a more reliable predictor of clustering behavior. For the ACP Network, we can see that the link type $\langle P, C \rangle$ has the weight 3.13, whereas the link type $\langle P, A \rangle$ has a much higher weight 13.30. This suggests that the latter link type is more reliable in deciding the cluster for papers, since a conference usually covers a broader spectrum than an author. For example, it is difficult to judge the cluster for a paper if we only know that it is published in the CIKM conference. The ability of our algorithm to learn such important characteristics of different link types is one of the reasons that it is superior to other competing methods.

For the weather sensor network, we summarize the link type strengths for the three networks with different network sizes that contain 5 observations for each sensor using the configuration of Setting 1, in Table 5. It is evident that *GenClus* correctly detects: (1) the P-typed sensors cannot be trusted as much as the other ones when P-typed sensors are very sparse, due to their farther distance and less similarity to other objects (the strengths of $\langle T, P \rangle$ and $\langle P, P \rangle$ relations decrease as $\#P$ decreases); and (2) for both types of sensors, T-typed neighbors are more trustable than P-typed ones, due to the higher quality of T-typed data in the network setting.

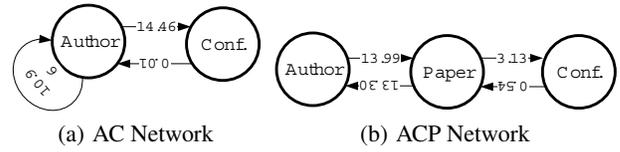

(a) AC Network    (b) ACP Network

**Figure 9: Strength for Link Types in Two Four-area Networks**

|   | $\langle T, T \rangle$ | $\langle T, P \rangle$ | $\langle P, T \rangle$ | $\langle P, P \rangle$ |
|---|---|---|---|---|
| T:1000; P: 250 | 3.14 | 2.88 | 1.60 | 1.32 |
| T:1000; P: 500 | 3.16 | 3.05 | 2.38 | 1.98 |
| T:1000; P: 1000 | 3.14 | 3.03 | 3.34 | 2.78 |

**Table 5: Link Type Strength for Weather Sensor Network in Setting 1**

## 5.3 A Typical Running Case

One of the core ideas of this paper is to enable a *mutual learning process* between the importance of link types for clustering and the actual clustering results. In this section, we provide some detailed results at different iterations of the algorithm, which suggests that such a mutual learning process does occur. In particular, a typical running case for the AC Network is illustrated in Fig. 10. Fig. 10(a) shows how the clustering accuracy progresses along with the changes in the importance of different link types. Fig. 10(b) shows how the strength weights change along with the clustering results at different iterations and finally converge to the correct values. Note that, we plotted the initial value $\boldsymbol{\gamma}$ at iteration 0 in Fig. 10(b), which is an all-one value.

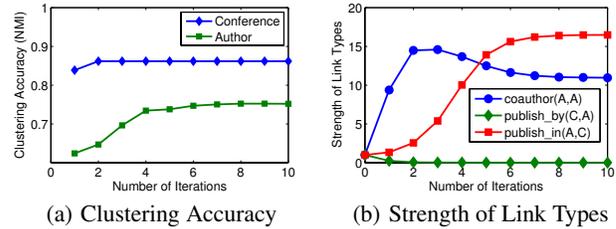

(a) Clustering Accuracy    (b) Strength of Link Types

**Figure 10: A Running Case on AC Network: Iterations 1 to 10**

## 5.4 Efficiency Study

To examine the efficiency of our algorithm, we illustrate the execution time of each inner iteration for the EM algorithm, which is the bottleneck component for the overall time complexity. The results are presented for the weather sensor network with different sizes and different numbers of observations for both pattern generator settings. The results are illustrated in Fig. 11, and are consistent with our observations in the complexity section about the scalability with the number of objects.

One observation is that the EM approach is very easy to parallelize, which is the major component for *GenClus*. We tested the parallel version of the EM algorithm with the use of 4 parallel threads (each running on a 2.13 GHz processor), and it turned out that the execution time is improved by a factor of 3.19. This suggests that the approach is highly parallelizable.

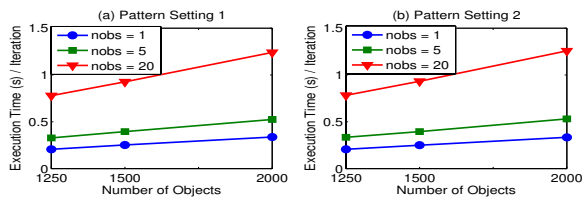

Figure 11: Scalability Test over Number of Objects

## 6. RELATED WORK

Clustering is a classical problem in data analysis, and has been studied extensively in the context of multi-dimensional data [13]. Most of these algorithms are attribute based, in which the data corresponds to a multi-dimensional format, and does not contain links. A number of clustering methods [5, 14, 6, 7] have been proposed on the basis of network structure only, mainly in the context of the community detection problem [2, 15, 8]. A recent piece of work extends the network clustering problem to the heterogeneous scenario [23]. However, this latter method [23] is designed for a specific kind of network structure, referred as the *star network schema*, and is not applicable to networks of general structure. Furthermore, it cannot be easily integrated with attribute information.

Recently, some studies [3, 20, 18, 25] have shown that by considering the link constraints in addition to the attributes, the clustering accuracy can be enhanced. However, most of these algorithms require that the network links, objects and their attributes are all homogeneous. A recent clustering method [28] integrates the network clustering process with categorical attributes by considering the latter as augmented objects, but the same methodology cannot be applied to numerical values. Some other algorithms [20] can cluster objects with numerical attributes by combining the network clustering objective function with a numerical clustering objective function, but it is difficult to decide the weight to combine them, and cannot deal with the incomplete attributes properly. [16] provides a framework for clustering objects in relational networks with attributes. However, they studied a different clustering problem by clustering objects from different types separately, and did not study the interplay of importance of different link types and the clustering results. Probabilistic relational models, such as [24], provide a way to model a rational database containing both attributes and links, but do not consider the scenario studied in this paper that clustering purposes could be different according to the specified attributes. Also, they cannot handle the problem of incomplete attributes due to the discriminative nature of their methods.

There are several different philosophies on using the link information in addition to attributes to help the clustering in networks. First, in [20, 28], links are viewed to provide another angle of similarity measure between objects besides the attribute-based similarity measure, and the final clustering results are generated by combining the two angles. Second, In relational clustering [16] and probabilistic relational models [24], every link is treated as equally important and the probability of a link appearance is modeled explicitly according to the cluster memberships of the two objects of the link, in a way of building mixture of block models [1]. Third, in [18, 22], links are considered to provide additional information about the similarity between objects that are consistent with the attributes, and the final clustering result is a more smoothing version compared with the one merely using attributes. However, none of these views is able to model the fact that different relations should have different importance in determining the clustering process for a certain purpose. Our philosophy in modeling link consistency is more similar to the third line, that is, two objects linking together indicates a higher chance that they have similar cluster memberships. Moreover, we further associate each type of links with a different importance weight in measuring the consistency under a given clustering purpose, and thus each type of relation carries different strengths in passing the cluster membership between the linked objects.

## 7. CONCLUSIONS

We propose *GenClus*, the first approach to cluster general heterogeneous information networks with different link types and different attribute types, such as numerical or text attributes, with guidance from a specified subset of the attributes. Our algorithm is designed to seamlessly work in the case when some of the nodes may not have the complete attribute information. One key observation of the work is that heterogeneous network clustering provides a tremendous challenge because different types of links may present different levels of semantic importance to the clustering process. The importance of different semantic link types is learned in order to enable an effective clustering algorithm that meets a user's demand. We present experimental results which show the advantages of the approach over competing methods, including a number of interesting case studies and a study of the algorithm efficiency.

# APPENDIX
## A. EM ALGORITHM PROOF

In the E-step of $t$th iteration, the $Q$ function, namely, the expected value of $g_1$ under the conditional distribution of hidden variables $Z^t$ with the meaning of cluster labels, given the observations $\{v[X]\}$ and current parameters $\Theta^{t-1}, \beta^{t-1}$, is:

$$Q = E_{Z^t|\{v[X]\}_{v \in V_X}, \Theta^{t-1}, \beta^{t-1}}(g_1(\Theta, \beta, Z^t))$$
$$= \sum_{Z^t} p(Z^t|\{v[X]\}_{v \in V_X}, \Theta^{t-1}, \beta^{t-1}) g_1(\Theta, \beta, Z^t)$$

where the link feature function $f$ and mixture model function in $g_1(\Theta, \beta, Z^t)$, the complete likelihood function of $g_1(\Theta, \beta)$, can be expanded by substituting with Eqs. (6) and (3) in Eq. (9):

$$g_1(\Theta, \beta, Z^t)$$
$$= \sum_{e=\langle v,u \rangle} \gamma(\phi(e))w(e) \sum_{k=1}^{K} \theta_{u,k} \log \theta_{v,k} + \sum_{v \in V_X} \sum_{l=1}^{m} c_{v,l}(\log \theta_{v,z} \beta_{z,l})$$

Since the feature function $f$ (contained in the first part of $g_1$) does not involve the observations of attributes and thus contains no hidden cluster label for each observation, the conditional expectation under $Z^t$ of $f$ is just $f$ itself. Therefore, the $Q$ function is then:

$$Q = \sum_{e=\langle v,u \rangle} \gamma(\phi(e))w(e) \sum_{k=1}^{K} \theta_{u,k} \log \theta_{v,k}$$
$$+ \sum_{v \in V_X} \sum_{k=1}^{K} \sum_{l=1}^{m} c_{v,l}(\log \theta_{v,k} \beta_{k,l}) p(z_{v,l}^t = k|\Theta^{t-1}, \beta^{t-1})$$

where the conditional probability for the hidden cluster label for object $v$ can be evaluated by: $p(z_{v,l}^t = k|\Theta^{t-1}, \beta^{t-1}) \propto \theta_{v,k}^{t-1} \beta_{k,l}^{t-1}$.

In the M-step, new values for parameters $\Theta^t$ and $\beta^t$ are achieved by maximizing the $Q$ function, with the help of introducing Lagrangian multipliers. First, the parameter $\theta_v^t$ for each object $v$ is maximized, by fixing the value of other parameters evaluated at step $t-1$, namely, $\{\theta_u^{(t-1)}\}_{u \neq v}$ and $\beta^{t-1}$, with the following updating rule for $k = 1$ to $K$:

$$\theta_{v,k}^t \propto \sum_{e=\langle v,u \rangle} \gamma(\phi(e))w(e)\theta_{u,k}^{t-1} + \mathbf{1}_{\{v \in V_X\}} \sum_{l=1}^{m} c_{v,l} p(z_{v,l}^t = k|\Theta^{t-1}, \beta^{t-1})$$

where $\mathbf{1}_{\{v \in V_X\}}$ is the indicator function, which equals to 1 if $v$ contains the attribute $X$, otherwise 0.

Then the parameter $\beta_k^t$ is evaluated by fixing $\Theta = \Theta^t$ for each cluster $k$, using the following updating rule for $l = 1$ to $m$: $\beta_{k,l}^t \propto \sum_{v \in V_X} c_{v,l} p(z_{v,l}^t = k|\Theta^{t-1}, \beta^{t-1})$.

## B. CONCAVITY PROOF

THEOREM 1. $g_2'(\gamma)$ *defined in Eq. (14) is a concave function.*

PROOF. To show $g_2'$ is a concave function, we only need to show $Hg_2'(\gamma)$ is a negative definite matrix, the $(i,j)$ element of which is

$$\frac{\partial g_2'(\gamma)}{\partial \gamma(r_i) \partial \gamma(r_j)} = \sum_{v=1}^{|V|} -\frac{1}{Z_v(\gamma)} \frac{\partial Z_v(\gamma)}{\partial \gamma(r_i) \partial \gamma(r_j)} - \frac{1}{\sigma^2} \mathbf{1}_{\{r_i = r_j\}}$$

where $Z_v(\gamma)$ is the normalization function for $p(\theta_v|\theta_{-v})$. Since each conditional distribution for $\theta_v$ belongs to the exponential family with parameters $\gamma$, then $\frac{\partial Z_v(\gamma)}{\partial \gamma(r_i) \partial \gamma(r_j)} = \mathbf{cov}_v(\gamma(r_i), \gamma(r_j))$, which is the covariance between $\gamma(r_i)$ and $\gamma(r_j)$. In all, $H(g_2')((\gamma)) = \sum_{v=1}^{|V|} -\frac{1}{Z_v(\gamma)} \mathbf{cov}_v - \frac{1}{\sigma^2} I$. Since for each object $v$, the corresponding covariance matrix $\mathbf{cov}_v$ is positive semidefinite, and the diagonal matrix denoted by $\frac{1}{\sigma^2} I$ is positive definite, then their linear combination with negative weights are negative definite. □

## C. SYNTHETIC WEATHER NETWORK GENERATOR

We now describe the weather sensor network generator. Assuming there are $K$ weather patterns, each of which is defined as a Gaussian distribution over temperature and precipitation attributes with different parameters. A weather sensor network is built by considering the sensors as the objects in the network, links denoting the $k$-nearest neighbors relationship, and temperature and precipitation as attributes. Each sensor is a mixture model of different weather patterns, and nearby sensors have similar pattern coefficients. Each sensor may have multiple observations, obtained at different times. The following specific steps and input parameters are required to enable the generation of the weather sensor network:

- Network size. The number of temperature sensors is denoted by $\#T$, the number of precipitation sensors by $\#P$, and the number of nearest neighbors required for link construction by $k$. These are input parameters to the generation process.

- Network structure. For each sensor, we randomly assign its location within a unit circle from the central point. An out-link exists between sensors $i$ and $j$, if $j$ is one of the $k$ nearest neighbors (of the particular type corresponding to $j$) from $i$.

- Weather pattern. Let $K$ be the number of clusters (weather patterns). Each such pattern is specified with a mean and covariance matrix over temperature and precipitation. The circle is then partitioned equally into $K$ rings, on the basis of distance from the central point.

- Cluster membership. The cluster membership for each sensor is determined by their reciprocal of the distance to the center for each weather region.

- Attribute observations. The number of observations is regulated by the user-specified input parameter $\#obs$. The attribute values at each sensor are generated according to the mixture model with the coefficients specified in its cluster membership.